\documentstyle[11pt]{article}

\def\<{\langle}
\def\>{\rangle}

\title{ENTROPY GENERATION IN COSMOLOGICAL PARTICLE CREATION}

\author{Mario A. CASTAGNINO,$^*$ \
 Fabi\'an H.    GAIOLI,$^{\dag}$ \ and Daniel M.
SFORZA$^{\ddag}$ \\ {\it Instituto de Astronom\'\i a y F\'\i sica del Espacio (IAFE),}
\\{\it Casilla de Correo 67, Sucursal 28, 1428 Buenos Aires, Argentina.}}

\vspace{1.0in}
\date{(Received}

\begin{document}

\baselineskip 1cm

\maketitle

\begin{abstract}

A very simplified model of the Universe is considered in order to propose  an alternative
approach to the  irreversible  evolution  of the Universe at very early times.  The
entropy generation at the quantum stage can be thought as a consequence of an instability
of the  system.   Then particle creation arises from this instability.

\end{abstract}

\vspace{3in}

\noindent $^*$e-mail: castagni@iafe.uba.ar

\noindent $^{\dag}$e-mail:    gaioli@iafe.uba.ar

\noindent $^{\ddag}$e-mail: sforza@iafe.uba.ar

\vfill\eject

Entropy  generation  in cosmological particle creation has  been  extensively studied
during the last years.  Essentially, the formalisms were developed in the framework of
quantum field theory in curved spacetime  and  at this level it    was  already
demonstrated  the  dissipative  nature  of  this  process \cite{hp,hk,k,ch}.    The notion
of entropy and its ``emergence'' in many  of these  descriptions    \cite{hk,k}  is  based
on  the  concept  of  ``coarse graining,'' by means  of  projection  operator  techniques.
This approach is also  successfully  realized  in    the    framework   of  quantum
cosmology \cite{roman}.  However, entropy generation  was really understood only in the
later stages of cosmic evolution.   In  the quantum creation regime, from the usual  point
of view, it was not  clear  which  mechanism  leads  to  entropy generation.  It is argued
that the evolution  of  the system follows strictly quantum  mechanical laws which are
``time-reversal invariant.'' Then, without external influence, we cannot expect any change
in entropy \cite{hp}.

An alternative point of view is to consider that quantum  mechanical laws are not
``always''  time-reversal  invariant.  There are many models for quantum unstable systems
(See, e.g. Ref. \cite{varios}); among them, a very  relevant case is the Friedrichs model
\cite{Frie}, which was widely  studied using the rigged Hilbert space formulation  of
quantum  mechanics \cite{varios2}.  This model  becomes  of  interest  to us because  our
cosmological  model can  be reduced to the Friedrichs one in a short time approximation,
as we shall see.

Let us introduce the model.   We  shall  use  a quantum, conformally coupled, massive
scalar field to represent the matter  degrees  of freedom.  Then, the action $S$ reads

\begin{equation}
S=\int d^4x \sqrt{-g}N\left\{{m_p^2\over 12}R-{1\over 2}\left[g^{\mu
\nu}\partial_{\mu}\Phi\partial_{\nu}\Phi+\xi R {\Phi}^2+{1\over
2}m\Phi^2\right]\right\},\label{acc1} \vspace{0.2in} \end{equation}

\noindent where $m$ stands for the mass of the fields quanta.

We  shall    also    assume    that   the  metric  corresponds  to  a  closed
Friedmann-Robertson-Walker  model.      Hence,   the  scalar  field  must  be homogeneous
in order to  satisfy  the  Einstein  equations.  Nevertheless, we want to go beyond the
two degrees of freedom of this minisuperspace and deal with  a  statistical  system. So,
to  obtain  a  consistent  model  with inhomogeneous matter fields, we shall not impose
the momenta constraints but only  the  Hamiltonian  constraint  averaged over each spatial
hipersurface. Formally, it is expressed in the fact that the  lapse  function $N$  depends
only on time (for a detailed discussion see Ref.  \cite{roman}).

Then the interval results

\begin{equation}
ds^2= -N^2(t) dt^2 + q^2(t) g_{ij} dx^i dx^j,\label{int} \vspace{0.2in} \end{equation}

\noindent where $g_{ij}$ is the induced metric on the spatial hipersurface labeled with
time $t$.  We can express an arbitrary scalar field configuration in terms of the
eigenfunctions of the spatial Laplacian, the spherical harmonics on  the 3-sphere
$Q^n_{lm}$:

\begin{equation}
\Phi(t,x)={1\over q}\sum_n \phi_n(t)Q_n(x),\label{campo} \vspace{0.2in} \end{equation}

\noindent where $n$ denotes the set $\{n,l,m\}$.

Therefore, we can derive the Hamiltonian

\begin{equation}
H={N\over 2q}\left\{-{1\over m_p^2}\pi^2 -m_p^2q^2 +\sum_n \left[p_n^2+ (n^2+ m^2 q^2)
\phi_n^2\right]\right\},\label{ham} \vspace{0.2in} \end{equation}

\noindent where $\pi=(-m_p^2/N)\dot{q}$ and $p_n=(q/N)\dot{\phi}_n$  are  the canonical
momenta associated with $q$ and $\phi_n$, respectively.

At this  point,  we  shall  not  follow  the  standard canonical quantization procedure.
In  order  to  recover a notion of time in quantum gravity (for a valuable treatise on
this  subject see Ref.  \cite{kuchar}), we can break the temporal reparametrization
invariance and choose  a privileged time.  In this framework, we can use the
``probabilistic  time,''  previously  introduced by one of us \cite{casta}.

We break  the temporal reparametrization invariance fixing the lapse function $N$.  Each
particular choice for the fixing implies a different probabilistic time.  In our  case,
the  more  appropriated  one  is  to  choose $N=q$, and $\theta$ turns out analogous to
the conformal time in quantum field theory in curved  spacetimes  (for  details  see Refs.
\cite{gordo,chile}).    Then, following the concepts of Ref.  \cite{fer},  we  can obtain
a Schr\"{o}dinger equation, where the temporal evolution is labeled by the probabilistic
time

\begin{equation}
i{\partial\Psi \over \partial{\theta}}=\left\{{1\over 2m_p^2}\partial_q^2 -{m_p^2\over
2}q^2 +{1\over 2} \sum_n \left[-\partial_{\phi_n}^2+ (n^2+ m^2 q^2)
\phi_n^2\right]\right\}\Psi.\label{ecscho} \vspace{0.2in} \end{equation}

The  Hamiltonian  can  be  rewritten  in  terms of  the  usual  creation  and annihilation
operators as

\begin{equation}
H=-\Omega_0 a^{\dagger}a+\int_0^{\infty}d\Omega \Omega
b^{\dagger}_{\Omega}b_{\Omega}+\lambda\int_0^{\infty}d\Omega
g(\Omega)(a+a^{\dagger})^2(b_{\Omega}+b^{\dagger}_{\Omega})^2~+ {\rm const},\label{hamc}
\vspace{0.2in} \end{equation}

\noindent where we have regularized the expression of $H$ and we have slightly modified it
without  substantially  affecting the qualitative behavior of the system. This  trick
allows  us to show the dissipative characteristic of the  system because the Poincar\'e
period becomes infinite \cite{chile,tasaki}.

It is very hard to deal with the interaction term.  However,  if we take some physical
processes  into  account,  we  can  obtain  a diagonalization of the Hamiltonian.  It  is
easy  to  verify  that  the  interaction  term does not conserve the particle number,  it
``connects''  states  which  differ  in  a multiple  of  pairs  of  particles.        For
giving  an  explicit  matrix representation,  we  adopt  the basis associated  to  the
``particle  number representation,'' since in the limit of weak particle creation (which
will be sufficient  to  demonstrate  the  existence  of  dissipation) the  number  of
particles is an adiabatic invariant.  In addition, we shall consider particle creation in
the first stages of the cosmic evolution, when occupation numbers are  small  and
spontaneous  particle  creation  dominates  over  stimulated creation \cite{hp}.   Then,
in  a  first approximation, at small times since Universe creation, we can  consider  the
creation  of a pair of particles in each mode, and only in  the  lowest  occupation number
states.  This sort of approximation is widely used in nuclear  physics, where a collective
state is obtained by diagonalizing the interaction in a  limited number of shell-model
states of particle excitation.  It is usual  to refer to such calculations as the
``Tamm-Dancoff method'' \cite{lane}.  Then, under these conditions we can consider only
the following states:

\vspace{0.2in}

\centerline{$    |1;0,...,0,...\>\equiv    |1\>$    \    \    and      \    \
$|1;...,2_{\Omega},...\>\equiv |\Omega \>$,}

\vspace{0.2in}

\noindent where the  first  place  corresponds to the quanta associated to $a$, and the
following places correspond  to  the  bath  (scalar  field)  modes.   So, the Hamiltonian
in this sector reads

\begin{equation}
H=\pmatrix{ -{\Omega}_0+3\lambda\int g(\Omega)d\Omega &\ldots&3\sqrt 2\lambda
g(\Omega)&\ldots  \cr  \vdots    &\ddots&    ~&~    \cr    3\sqrt    2\lambda
g(\Omega)&~&-{\Omega}_0+2\Omega + 12\lambda            g(\Omega)+3\lambda\int
g(\Omega)d\Omega&~  \cr  \vdots&~&~&\ddots\cr}.\label{mcosmo} \vspace{0.3in}
\end{equation}

Redefining

\begin{equation}
\omega_0=-{\Omega}_0+3\lambda\int_0^{\infty} g(\Omega)d\Omega,\label{def1}
\end{equation}
\begin{equation}
\omega=\omega_0+2\Omega,\label{def2} \vspace{0.2in} \end{equation}

\noindent we  obtain  the  Hamiltonian    of  a  Friedrichs  model  for  this  subspace
(one-particle sector of an oscillator  in  a bosonic reservoir \cite{gordo}), in the form

\begin{equation}
H=\pmatrix{  \omega_0&\ldots&  \frac{3\sqrt    2}{2}\lambda   g(\frac{\omega-
\omega_0}{2}) &\ldots \cr \vdots  &\ddots&  ~&~ \cr \frac{3\sqrt 2}{2}\lambda
g(\frac{\omega- \omega_0}{2}) &~&\frac{\omega}{2} + 6\lambda  g(\frac{\omega-
\omega_0}{2})&~ \cr \vdots&~&~&\ddots\cr}.\label{mcos} \vspace{0.3in} \end{equation}

\noindent It can be rewritten as

\begin{equation}
H=\omega_0  |1\>\<1| +  \int_{\omega_0}^{\infty}d\omega \left[\frac{\omega}{2}  + 6\lambda
g\left(\frac{\omega-    \omega_0}{2}\right)\right] |\omega\>\<\omega|        +
\lambda\int_{\omega_0}^{\infty}d\omega  \frac{3\sqrt{2}}{2} ~g\left(\frac{\omega -
\omega_0}{2}\right)(|1\>\<\omega| + |\omega\>\<1|),\label{frie1} \vspace{0.2in}
\end{equation}

\noindent where ${\omega}_0~\in~{\cal  R}_{\geq  0}$~~and  $g(0)=0$.    The  associated
Hilbert      space    is        ${\cal        H}={\cal        C}~\oplus~{\cal
L}^2(\frac{\omega_0}{2},\infty)$.

The  eigenvalue  problem  can  be exactly solved (see Ref.   \cite{varios2}). This
well-known  solution    includes    eigendistributions  which  must  be interpreted in an
adequate extension of the Hilbert space.

In order to make  explicit  an  exponential  decaying law we can analytically extend the
problem to the  complex plane (for a detailed derivation see Refs. \cite{varios2,gordo}).
In this case, the  second-sheet  extension  of  the reduced resolvent of the Hamiltonian,

$$\alpha^{-1}(z)= \left[z-{\omega_0}- \frac{9}{2}{\lambda}^2 \int_{\omega_0}^{\infty}
d\omega {g^2(\frac{\omega  -\omega_0}{2})  \over z-\frac{\omega}{2}- 6\lambda
g(\frac{\omega- \omega_0}{2})}\right]^{-1},$$

\vspace{0.2in}

\noindent presents a complex pole $z_0$.  Thus, in a generalized spectral decomposition of
the Hamiltonian $z_0$ appears as a ``complex  eigenvalue'' \cite{varios2}. For small
values of $\lambda$, $z_0$ can be estimated as \cite{bnhh}

\begin{equation}
z_o\approx  \left[  \omega_0  +  {\cal  P}
\int_{\omega_0}^{\infty}d\omega~{\frac{9}{2}{\lambda}^2   g^2(\frac{\omega  -
\omega_0}{2})  \over  \omega_0-  \frac{\omega}{2}-  6\lambda  g(\frac{\omega-
\omega_0}{2})}  \right]  +  i\left[-\pi  \frac{9}{2}{\lambda}^2 g^2(\frac{\omega_0}{2})
\right],\label{z} \vspace{0.2in} \end{equation}

\noindent where ${\cal P}$ denotes principal part.  $z_0$ is usually expressed as

\begin{equation}
z_0=\tilde{\omega}_0 - i{\gamma \over 2} \ \ \ (\gamma>0),\label{z2} \vspace{0.2in}
\end{equation}

\noindent where  $\tilde{\omega}_0-\omega_0$  is  the level shift and $1/\gamma$ is the
mean-life of the unstable level.

By performing  complex  integration and using the residues theorem, a complex contribution
coming from  the  pole $z_0$ gives rise to a set of states which we shall use to  obtain a
spectral  resolution  of  $H$.    The  associated eigendistributions (corresponding to the
discrete  part) from right and left, respectively are \cite{varios2}

\begin{eqnarray}
&|1^-\>=[\alpha_{II+}^{\prime}(z_0)]^{-1/2}\left(|1\> + \int_{\Gamma}  d\omega
{\frac{3\sqrt{2}}{2}  \lambda  g(\frac{\omega  - \omega_0}{2})\over{z_0  -
\frac{\omega}{2} -6\lambda    g(\frac{\omega -\omega_0}{2})}}|\omega\> \right),\nonumber
\\ & \\ &\<1^+|=[\alpha_{II+}^{\prime}(z_0)]^{-1/2}\left(\<1| + \int_{\Gamma} d\omega
{\frac{3\sqrt{2}}{2}  \lambda  g(\frac{\omega -  \omega_0}{2})\over{z_0  -\frac{\omega}{2}
-6\lambda g(\frac{\omega   -\omega_0}{2})}}\<\omega| \right),\nonumber \label{4.6.51}
\end{eqnarray}

\vspace{0.2in}

\noindent where the curve $\Gamma$ passes just under the pole $z_0$ and $II$ stands for
the  second  Riemann-sheet.    The vectors corresponding  to  the  continuous solution are

\begin{eqnarray}
&|\omega^-\>=|\omega\>     +    \frac{3\sqrt{2}}{2}{\lambda    g(\frac{\omega
-\omega_0}{2})\over{\tilde\alpha_{+}(\omega)}}\left[|1\>          +
\int_{\omega_0}^{\infty}    d\omega^{\prime}      {\frac{3\sqrt{2}}{2}\lambda
g(\frac{\omega^{\prime} -\omega_0}{2})\over{\omega -\frac{\omega^{\prime}}{2} -6\lambda
g(\frac{\omega^{\prime}        -\omega_0}{2})        +
i\epsilon}}|\omega^{\prime}\>\right],\nonumber \\ & \\ &\<\omega^+|=\<\omega|     +
\frac{3\sqrt{2}}{2}{\lambda    g(\frac{\omega
-\omega_0}{2})\over{\alpha_-(\omega)}}\left[\<1|  + \int_{\omega_0}^{\infty}
d\omega^{\prime}       {\frac{3\sqrt{2}}{2}\lambda    g(\frac{\omega^{\prime}
-\omega_0}{2})\over{\omega        -\frac{\omega^{\prime}}{2}        -6\lambda
g(\frac{\omega^{\prime}                -\omega_0}{2})                       -
i\epsilon}}\<\omega^{\prime}|\right],\nonumber \label{4.6.52}
\end{eqnarray}

\vspace{0.2in}

\noindent where ${\alpha}_{\pm}=\alpha(\omega \pm i\epsilon)$ and

$$\frac{1}{\tilde\alpha_+(\omega)}  =  \frac{1}{\alpha_+(\omega)}  +  2\pi  i
\frac{\delta(\frac{\omega}{2}  +6\lambda  g(\frac{\omega  -  \omega_0}{2})  -
z_0)}{\alpha_{II+}^{\prime}(z_0)}.$$

\vspace{0.2in}

The orthonormality  and completeness relations read

\begin{equation}
\<1^+|1^-\>=1,~~~\<1^+|\omega^-\>=\<\omega^+|1^-\>=0,~~~ \<\omega^+|\omega^{\prime
-}\>=\delta(\omega-\omega^{\prime}), \label{4.6.53}
\end{equation}
\begin{equation}
|1^-\>\<1^+|        +       \int_{\frac{\omega_0}{2}}^{\infty}        d\omega
|\omega^-\>\<\omega^+| = 1. \label{4.6.55} \vspace{0.2in} \end{equation}

Finally, the spectral decomposition of $H$ is

\begin{equation}
H=z_0|1^-\>\<1^+|    +  \int_{\frac{\omega_0}{2}}^{\infty}  d\omega    \omega
|\omega^-\>\<\omega^+|. \label{4.6.54} \vspace{0.2in} \end{equation}

The eigenvectors (\ref{4.6.55}) (from right and left, respectively) and their associated
complex eigenvalue $z_0$  have  not  sense in the standard Hilbert space formulation of
quantum mechanics.    However, $H$ admits a self-adjoint extension to a ``rigged Hilbert
space,''  where  the  states corresponding to complex eigenvalues acquire meaning
\cite{varios2,BGM,gordo}.  The idea is to restrict the topology of the Hilbert space
${\cal  H}$  in  order  to  have a nuclear space $\Omega$ such that $\Omega\subset{\cal
H}$.  Then,  the  linear functionals  on  $\Omega$  belong  to  a  bigger space
$\Omega^{\times}$, the topological  dual  of  $\Omega$,  such  that  we build up a
Gel'fand  triplet \cite{Ge}:  \  \  $\Omega\subset{\cal H}\subset\Omega^{\times}$.  The
nuclear space is chosen by  using  convergence  prescriptions.    In  our  case, this
condition is fulfilled if we  use Hardy class functions \cite{PW,Ga} in order to define
two nuclear subspaces $\Omega_+$  and $\Omega_-$, for the upper and the  lower half-plane,
respectively (Hardy classes from  above  and  below, respectively) \cite{laura}.  From the
Paley-Wiener theorem \cite{PW}  we know that the time evolution splits into two
semigroups, i.e.

\begin{eqnarray}
&|1^-(\theta)\>= e^{-iH\theta}|1^-\>= e^{-i\tilde{\omega}_0\theta}e^{-{\gamma\over
2}\theta}|1^-\>,~~~~~  \theta>0, \nonumber \\ & \\ &|1^+(\theta)\>= e^{-iH\theta}|1^+\>=
e^{-i\tilde{\omega}_0\theta}e^{{\gamma\over 2}\theta}|1^+\>,~~~~~  \theta<0. \nonumber
\label{evtemp}
\end{eqnarray}

\vspace{0.2in}

Then, the state  $|1^-(\theta)\>$  decays  towards  the  past  and  the state
$|1^+(\theta)\>$ decays towards the future.  These vectors belong to the dual spaces  of
$\Omega_+$ and $\Omega_-$,  respectively  $\Omega_+^{\times}$  and $\Omega_-^{\times}$.
The    physical    meaning    of        the $\Omega_{\pm}^{\times}$-states is the
following:  the state  $|1^-(\theta)\>$ corresponds to the evolution from an unstable
state  towards  a  stable  one. The  state  $|1^+(\theta)\>$  corresponds to the formation
process (which  is usually neglected in the case of a large mean-life).

The survival amplitude of the state $|1\>$ in time $\theta$ is given by

\begin{equation}
\<1|e^{-iH\theta}|1\>={1\over{\alpha_{II+}^{\prime}(z_0)}}e^{-iz_0\theta}   +
\frac{9}{2}\lambda^2\int^{\infty}_{\omega_0}                      e^{-i\omega
\theta}{g^2(\frac{\omega                       -                \omega_0}{2})
\over{\alpha_+(\omega)\alpha_-(\omega)}}d\omega. \label{toto} \vspace{0.2in}
\end{equation}

The  first  term  is  dominant  over  the second for short  times  (with  the exception of
the  short  initial  {\it  Zeno} period \cite{MS}), showing an exponential  behavior. On
the other hand, the second term is the  continuous background  and  is  relevant  for
later  times.    In  this way the initial matter-less state  $|1\>$  is  unstable,
favouring transitions to states with greater number of particles with a smooth exponential
law.

Formally, this dissipative  behavior  can  be  made explicit by the following
``H-theorem.'' Let us consider the operator \cite{found}

\begin{equation}
{\cal Y}=|1^-\>\<1^-|.\label{lyapu} \vspace{0.2in} \end{equation}

\noindent It evolves according to the Heisenberg equation of motion

\begin{equation}
{\cal                                         Y}(\theta)=e^{iH\theta}{\cal
Y}(0)e^{-iH\theta}=e^{\gamma\theta}{\cal Y}(0).\label{lyapu2} \vspace{0.2in}
\end{equation}

\noindent ${\cal Y}(\theta)$ is a monotonic increasing function in time $\theta$, i.e.

\begin{equation}
{d{\cal  Y}\over d{\theta}}>0.\label{deriv} \vspace{0.2in} \end{equation}

\noindent Any component of this operator  is  therefore a Lyapunov function.  Hence, in
this model, ``dissipation'' results as a consequence of the resonance between the scale
factor and the scalar field (this fact was already suggested as the responsible  of  an
instability at the classical level  with  the  consequent apparition  of chaos
\cite{estcla}), promoting states with greater occupation numbers,  as  is  usually
expected.  Thus, dissipation can be  thought  as  a generalized  particle  creation
process,  the  vacuum of the matter field is unstable  under  pair  creation, whereby
changes in the ``system'' (i.e.  the degree  of  freedom  related  to  gravitation,  the
scale  factor)  lead  to excitations of degrees of freedom of the ``bath'' (i.e.  the
infinity degrees of freedom of the  scalar matter field), which then propagate or diffuse
away \cite{ch}.  What we have found is that these excitations are originated in an
intrinsic resonance (associated with the pole  $z_0$)  between the system and the bath.

Although the model we have presented here  cannot  be  considered  as  a very realistic
one, it constitutes a first step for an alternative approach to the problem of finding the
yet ``unclear'' mechanism of dissipation  at the early stage of the Universe.

\vspace{2in}

\section*{Acknowledgements}

This work was partially supported by Universidad de Buenos Aires.

\vfill\eject

\end{document}